\documentclass[12pt,preprint]{aastex}
\usepackage{lscape}
\usepackage{multirow}
\usepackage{booktabs}
\usepackage{graphicx}
\usepackage{amsmath}
\usepackage{longtable}
\usepackage{rotating}

\newcommand{\sigmagas}{\Sigma_{\rm gas}}
\newcommand{\msun}{M_\odot}
\newcommand{\mdot}{\dot{M_\star}}
\newcommand{\mdisk}{M_{\rm disk}}
\newcommand{\mstar}{M_\star}
\newcommand{\omegap}{\Omega_{\rm p}}

\begin{document}
\title{Spiral Arms in Gravitationally Unstable Protoplanetary Disks as Imaged in Scattered Light}

\shorttitle{Spiral Arms}

\shortauthors{}

\author{Ruobing Dong\altaffilmark{1,2,4}, Cassandra Hall\altaffilmark{3}, Ken Rice\altaffilmark{3}, Eugene Chiang\altaffilmark{2}}

\altaffiltext{1}{Nuclear Science Division, Lawrence Berkeley National Lab, Berkeley, CA 94720, rdong2013@berkeley.edu}
\altaffiltext{2}{Department of Astronomy, University of California at Berkeley, Berkeley, CA 94720}
\altaffiltext{3}{Institute for Astronomy, University of Edinburgh, Blackford Hill, Edinburgh EH9 3HJ, UK}
\altaffiltext{4}{NASA Hubble Fellow}

\clearpage

\begin{abstract}

Combining 3D smoothed-particle hydrodynamics and Monte Carlo radiative transfer calculations, we examine the morphology of spiral density waves induced by gravitational instability (GI) in protoplanetary disks, as they would appear in direct images at near-infrared (NIR) wavelengths. We find that systems with disk-to-star-mass ratios $q=\mdisk/\mstar$ that are $\sim$0.25 or more may produce prominent spiral arms in NIR imaging, remarkably resembling features observed in the MWC 758 and SAO 206462 systems. The contrast of GI-induced arms at NIR wavelengths can reach a factor of $\sim$3, and their pitch angles are about $10^\circ-15^\circ$. The dominant azimuthal wavenumber of GI-induced spiral arms roughly obeys $m\sim1/q$ in the range $2\lesssim1/q\lesssim8$. In particular, a massive disk with $q\approx0.5$ can exhibit grand-design $m=2$ spirals. GI-induced arms are in approximate corotation with the local disk, and may therefore trap dust particles by pressure drag. Although GI can produce NIR spiral arms with morphologies, contrasts, and pitch angles similar to those reported in recent observations, it also makes other demands that may or may not be satisfied in any given system. A GI origin requires that the spirals be relatively compact, on scales $\lesssim$ 100~AU; that the disk be massive, $q\gtrsim0.25$; and that the accretion rate $\mdot$ be high, on the order of $10^{-6}\msun$ yr$^{-1}$.

\end{abstract}



\section{Introduction}\label{sec:intro}

In the past few years, direct imaging at near-infrared (NIR) wavelengths has revealed a panoply of fine features in gaseous protoplanetary disks, thanks to the commissioning of several major high contrast, high angular resolution, adaptive optics imaging instruments onboard 8-meter class telescopes (e.g., Subaru/HiCIAO, \citealt{suzuki10}; VLT/SPHERE, \citealt{beuzit08}; Gemini/GPI, \citealt{macintosh08}). These observations image starlight scattered by dust grains in disk surface layers, and they take advantage of polarized intensity (PI) imaging \citep[e.g.,][]{perrin04,hinkley09} to achieve inner working angles as small as $0.1\arcsec$ (corresponding to $\sim$14 AU at the distance of Taurus) with sub-$0.1\arcsec$ angular resolution. The various kinds of radial and azimuthal structures revealed promise to shed light on planet formation processes in disks.


One particularly interesting discovery from NIR disk imaging surveys is the large scale spiral-arm-like features identified in 2 protoplanetary disks: MWC~758 \citep[][see also \citealt{grady13}]{benisty15} and SAO~206462 \citep[][see also \citealt{muto12}]{garufi13}. The primary in both cases is a Herbig star (1.8~$\msun$ in MWC 758, \citealt{chapillon08}; and 1.6~$\msun$ in SAO 206462, \citealt{andrews11}), and both systems are recognized as transitional disks with a giant depleted central region \citep{andrews11,perez14,marino15-mwc758}. The 2 major arms in each system are in nearly $m=2$ rotational symmetry, and are located at tens of to $\sim$100 AU from the center. The arms can be easily traced over at least 180$^\circ$ in the azimuthal direction, and they are quite open with large pitch angles (defined as the angle between the azimuthal direction and the tangent of the arm). Besides MWC 758 and SAO 206462, spiral arms on smaller azimuthal (and/or radial) scale have been discovered in a number of systems, such as AB Aur \citep{hashimoto11}, HD 142527 \citep{fukagawa06,canovas13,avenhaus14}, and HD 100546 \citep{currie14}. The number of arms in this second class of system is generally more than 5 (with the exception of HD 100546, which exhibits a couple of small arms), and the arms have an azimuthal extension generally no bigger than $90^\circ$.

Recently, \citet{dong15-spiralarms} have shown that $m=2$ arms such as the pair in MWC 758 may be explained by density waves excited by one giant companion (planet or brown dwarf) on the periphery of the arms (see also \citealt{zhu15-densitywaves}). In this paper, we investigate another physical mechanism that is also able to excite density waves in disks --- gravitational instability (GI; e.g., \citealt{rice03,lodato04, lodato05}). We focus on the observational signatures of GI-induced density waves in NIR imaging observations, as predicted by 3D smoothed-particle hydrodynamics (SPH) + Monte Carlo radiative transfer (MCRT) simulations. We ask specifically whether GI-induced features can resemble the observed spiral arms. This general topic has been explored by \citet{dipierro15-gidisks}, who studied the evolution of dust in a fixed gas disk background obtained from the \citet{lodato04} SPH GI simulations. They calculated 2D dust surface density distributions, which were then puffed up in the vertical direction by a parametrized gaussian profile and fed into a 3D Monte Carlo radiative transfer (MCRT) code to produce synthetic images.
These authors focused on relatively
low mass disks, computing 
spatial distributions of grains 
with different size distributions
and simulating both NIR scattered light
and mm-wave ALMA images (see also
an earlier paper by \citealt{dipierro14}
that treated gas and dust as well-mixed
and that concentrated on ALMA
observations).
By comparison, we will focus on NIR imaging, and will explore a larger range of disk masses, making direct comparisons between models and 
scattered light observations of
real astronomical sources.
\citet{pohl15} combined 2D hydro and 3D MCRT simulations to study the appearance of spiral arms at NIR wavelengths in GI disks with the added complication of an embedded planet. Our aim here is to isolate the effects of GI so that we can compare
planet vs.~GI-based interpretations
of NIR images.

The structure of our paper is as follows. The hydro and MCRT simulations are introduced in Section~\ref{sec:setup}; the modeling results are presented in Section~\ref{sec:results}; and a summary and connections to observations are given in Section~\ref{sec:summary}.


\section{SPH and MCRT Simulations}\label{sec:setup}

The general workflow of our modeling is as follows. We carry out 3D global SPH simulations to calculate the gas density structures of gravitationally unstable disks (Section~\ref{sec:hydro}). The resulting models are read into the \citet{whitney13} 3D dust MCRT code to produce synthetic model images at $1.6~\micron$ ($H$~band, Section~\ref{sec:mcrt}), assuming that dust is uniformly mixed with gas. The vertical structure of the SPH disk is iterated in the MCRT code until radiative equilibrium and hydrostatic equilibrium (HSEQ) are achieved. Finally, synthetic images are convolved by a Gaussian point spread function (PSF) to achieve an angular resolution comparable to recent observations.

\subsection{Smoothed-Particle Hydrodynamics Simulations}\label{sec:hydro}

The SPH simulations are based on the code developed by \citet{bate95}, augmented by a hybrid radiative transfer formalism that combines the polytropic approximation with the flux-limited diffusion approximation \citep{forgan09, stamatellos09} to model both global cooling and radiative transfer. The opacity and temperature of the gas are calculated using an equation of state that accounts for the effects of H$_2$ dissociation, H$^0$ ionization, He$^0$ and He$^+$ ionization, ice evaporation, dust sublimation, molecular absorption, bound-free and free-free transitions and electron scattering \citep{bell94,boley07,stamatellos07}. The simulations include heating of the disk through $P$d$V$ work and shocks; stellar irradiation is not included.

The mass of the central central star $M_\star$ is assumed to be 1~$\msun$. In total we construct 3 models (Table~\ref{tab:models}) where the disk-to-star-mass ratio $q=\mdisk/\mstar$ equals $0.125$, $0.25$ and 0.5, respectively. Each gas disk comprises $2 \times 10^6$ SPH particles, initially located between 10 and 60 AU, and distributed in such a way that the initial surface density profile $\Sigma_0 \propto  r^{-3/2}$ and the initial sound speed profile $c_{\rm s}\propto r^{-1/4}$. By limiting the initial disk to be within 60 AU, we ensure that these models remain quasi-steady, rather than fragmenting into bound companions \citep[e.g.,][]{rafikov05, stamatellos07,clarke09, rice09}. The accretion rate of the disks ($\mdot$ in the table) is calculated by counting the rate at
which particles accrete onto the star. All simulations are run for about 3500 years, or $\sim$10 orbits at 50 AU where the spiral arms are most prominent.

\subsection{Monte Carlo Radiative Transfer Simulations}\label{sec:mcrt}

Density structures obtained in our SPH simulations are post-processed using the Whitney MCRT code. We construct a 3D disk in spherical coordinates, with $400\times200\times256$ cells in $r$ (radial, from 5 to 100 AU with logarithmic spacing to better resolve the inner disk), $\theta$ (polar, from 0 to $\pi$ with power law spacing to better resolve the disk midplane at $\theta=\pi/2$), and $\phi$ (azimuthal, from 0 to $2\pi$ with uniform spacing).  Photons from an $R_\star=2$~$R_\odot$ and $T_\star=4500$~K central star (corresponding to a $1\msun$ star
that is 2 Myr old on the Hayashi track; \citealt{hayashi61}) are absorbed/reemitted or scattered by dust in the surrounding disk.\footnote{We also experimented with a 1.8$\msun$ pre-main-sequence star (as in the MWC 758 system; \citealt{chapillon08}) in the MCRT simulations, and found that the resulting scattered light images merely
brightened by a factor of $\sim$4 
because of the higher stellar luminosity.} The temperature in each grid cell is calculated based on the radiative equilibrium algorithm described in \citet{lucy99}. The power released from accretion $\mdot$ at each radius is deposited locally into the cells, and the accretion power between the magnetic truncation radius (assumed to be at $5R_\star$) and the stellar surface is added to the total luminosity of the star. The scattering phase function is approximated using the \citet{henyey41} function. Polarization is calculated assuming a Rayleigh-like phase function for the linear polarization \citep{white79}. All simulations are run with 1 billion photon packets.

Our MCRT simulations are run with the ``HSEQ'' switch in the Whitney code turned on. In some of our SPH simulations, the temperature in the outer disk reaches a minimum floor (artificially set in the SPH code) that is lower than what is estimated using the MCRT calculations. As mentioned in Section~\ref{sec:hydro}, we do not have stellar irradiation in the SPH simulations. As a result, the outer regions of our SPH disks can be artificially cold, and thin in the vertical direction. To correct this effect, we readjust the vertical density distribution of the disks in the MCRT simulations using the MCRT temperature and not the SPH temperature, while still using the surface density distribution from the SPH simulations. This inconsistency of method is not serious; if we were to use a higher temperature in the SPH simulations, the spiral structures would be only slightly weaker than suggested here \citep{rice11, rafikov15} for fixed disk mass;
thus the $q$-values reported here
are slightly underestimated.

In MCRT calculations, the vertical structure is solved based on a self-consistent HSEQ condition, as gas pressure balances gravity in the vertical direction $z$:
\begin{equation}
\frac{dp}{dz}=-\rho \frac{z}{r} \frac{GM(r)}{r^2},
\label{eq:hseq}
\end{equation}
where $p$ is pressure, $\rho$ is density, and $M(r)=M_\star+\mdisk(r)$ with $\mdisk(r)$ equal to the total disk mass within $r$. The quantities $T$ and $\rho$ are iterated until convergence is achieved. 

The dust grains in the disk are assumed to be interstellar medium (ISM) grains \citep{kim94} made of silicate, graphite, and amorphous carbon. Their size distribution is a smooth power law in the range of 0.02-0.25~$\micron$ followed by an exponential cut off beyond 0.25~$\micron$. These grains are small enough to be dynamically well coupled to the gas, so that their volume density $\rho_{\rm dust}$ is linearly proportional to the gas $\rho_{\rm gas}$; we take the ratio to be $\rho_{\rm gas}$:$\rho_{\rm dust}$=1000:1.\footnote{This corresponds to a conventional 100:1 gas-to-solid mass ratio and a 10:1 solid-to-ISM-dust mass ratio. The remaining 90\% of the solids are assumed to be in the form of large grains and/or planetesimals that may have settled to the disk midplane and do not affect NIR scattering.} The optical properties of the grains can be found in Figure 2 of \citet{dong12cavity}. 

Full resolution synthesized PI images at $H$~band ($1.6\micron$) are produced from MCRT simulations,\footnote{In this work, the physical quantity recorded in all model images is the specific intensity in units of
[mJy~arcsec$^{-2}$], or [$10^{-26}$ ergs~s$^{-1}$~cm$^{-2}$~Hz$^{-1}$~arcsec$^{-2}$].} and then convolved by a Gaussian PSF with a full width half maximum (FWHM) of $0.05\arcsec$ assuming a distance of 140~pc, to achieve an angular resolution comparable to those obtained with Subaru, VLT, and Gemini (the FWHM of a theoretical Airy disk is $1.028\lambda/D\sim0.04\arcsec$ for a primary mirror with a diameter $D=8.2$~m at $\lambda=1.6~\micron$).


\section{Results}\label{sec:results}

\begin{figure}
\begin{center}
\epsscale{0.5} \plotone{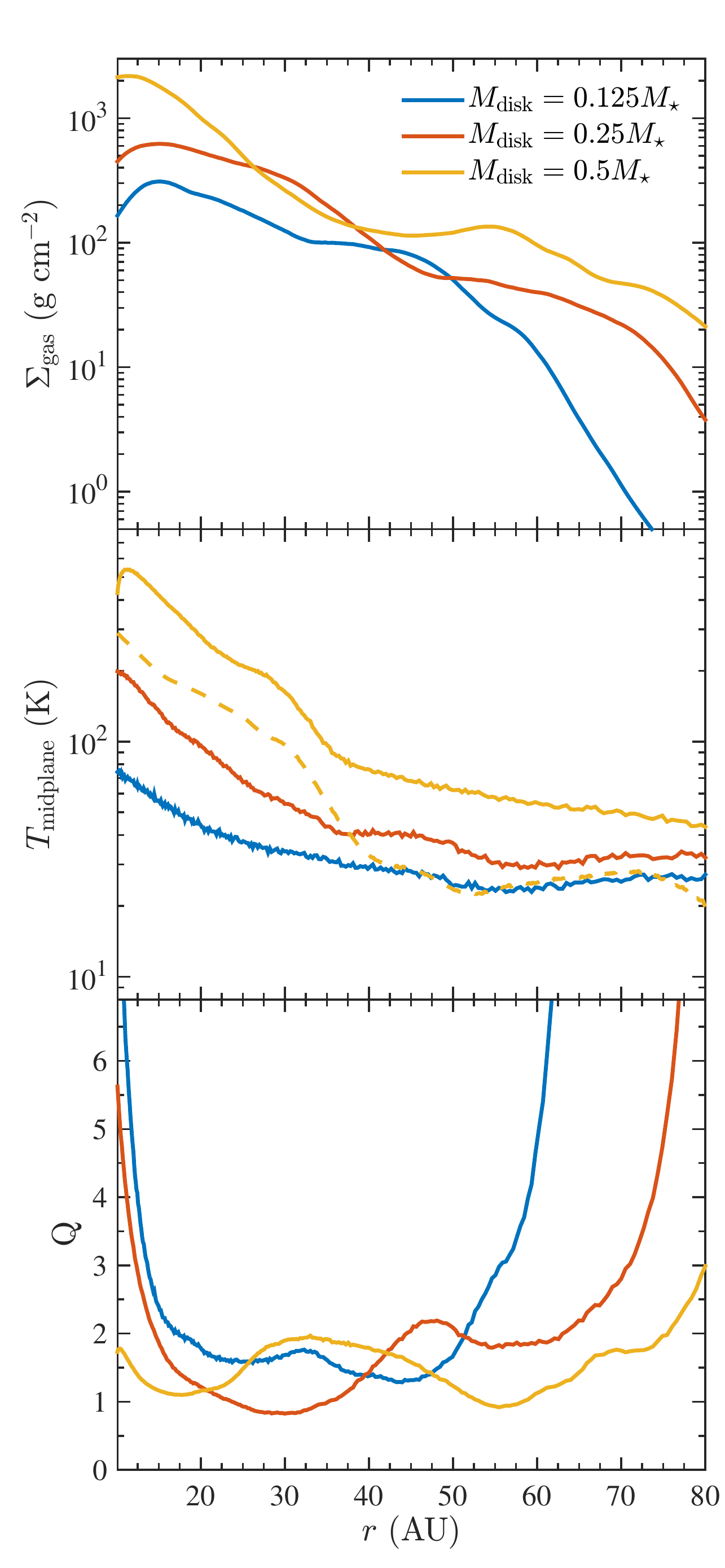}
\end{center}
\figcaption{Surface density (top, from SPH simulations), midplane temperature (middle, from MCRT simulations), and Toomre $Q$ (bottom, eqn.~\ref{eq:q}) for our models of various disk masses. The dashed curve in the middle panel is the midplane temperature from the SPH simulation for the $\mdisk=0.5\mstar$ model. The MCRT temperature can be up to a factor of $\sim2$ higher than the SPH temperature in the outer disk. 
\label{fig:hydro}}
\end{figure}

The radial profiles of the surface density, midplane temperature, and Toomre $Q$ for all models are shown in Figure~\ref{fig:hydro}. GI develops in all cases, and manifests itself in the form of spiral density waves. Disks become marginally gravitationally unstable when 
\begin{equation}
Q=\frac{c_{\rm s}\kappa}{\pi {\rm G} \Sigma_{\rm gas}}
\label{eq:q}
\end{equation}
is $\lesssim2$, where $\kappa$ is the epicyclic frequency. As $q$ increases, the $Q\lesssim2$ regime expands, from $\sim$20-50 AU in MD0125, to $\sim$20-60 AU in MD025, and to $\sim$10-70 AU in MD05.

As described in \citet{lodato05}, GI disks in our models ($q\leq0.5$) undergo 3 phases: (1) the initial settling phase that lasts for about 500 years, during which the disk adjusts its outer radius, and material
redistributes radially by axisymmetric evolution and ring formation; (2) the `burst' phase that lasts for about 1000 years, during which spiral density waves form and grow; and (3) the asymptotic phase, where GI is self-regulated and the disk reaches a quasi-equilibrium state. In the latter two phases, the dominant azimuthal wavenumber $m$ (Table~\ref{tab:models}) decreases as $q$ increases, consistent with \citet{lodato04} and \citet{cossins09}. In our $q$ range, we find $m\sim1/q$, with the approximation becoming less accurate at low $q$ values ($m$ does not increase as fast as $1/q$).

The spiral density waves have a pattern speed $\omegap$, listed in Table~\ref{tab:models} along with the corresponding corotation radius $r_{\rm CR}$ ($\Omega_{\rm Keplerian}(r_{\rm CR})=\omegap$). In general, the spirals are roughly in Keplerian rotation, and $r_{\rm CR}$ coincides with the main part of the arms. This is consistent with \citet{cossins09}, who found for a $q=0.125$ disk that $\omegap$ differs from the local $\Omega_{\rm Keplerian}$ by $\lesssim15\%$. This property should lead to dust trapping by the arms and promote grain growth and planetesimal formation. As shown by \citet{rice06} and \citet{paardekooper06}, dust particles with Stokes number $\sim$1 tend to be trapped at local pressure maxima in disks; this has been demonstrated both in radial \citep[e.g., ][]{pinilla12-diffcavsize,zhu12,dejuanovelar13,dong15-gaps} and azimuthal gas structures \citep[e.g., ][]{lyra13, zhu14votices, mittal15} as a mechanism to explain observed concentrations of dust particles in ALMA disks \citep[e.g., ][]{fukagawa13, casassus13, vandermarel13, perez14, vandermarel15}. In the case of spiral arms, \citet{rice04} have shown that for $q=0.25$, spiral density waves can act to trap dust particles, and through 2D shearing box simulations, \citet{gibbons12} concluded that the local dust surface density can be enhanced by up to $\sim3.5$ orders of magnitude relative to the gas.

\begin{figure}
\begin{center}
\epsscale{0.8} \plotone{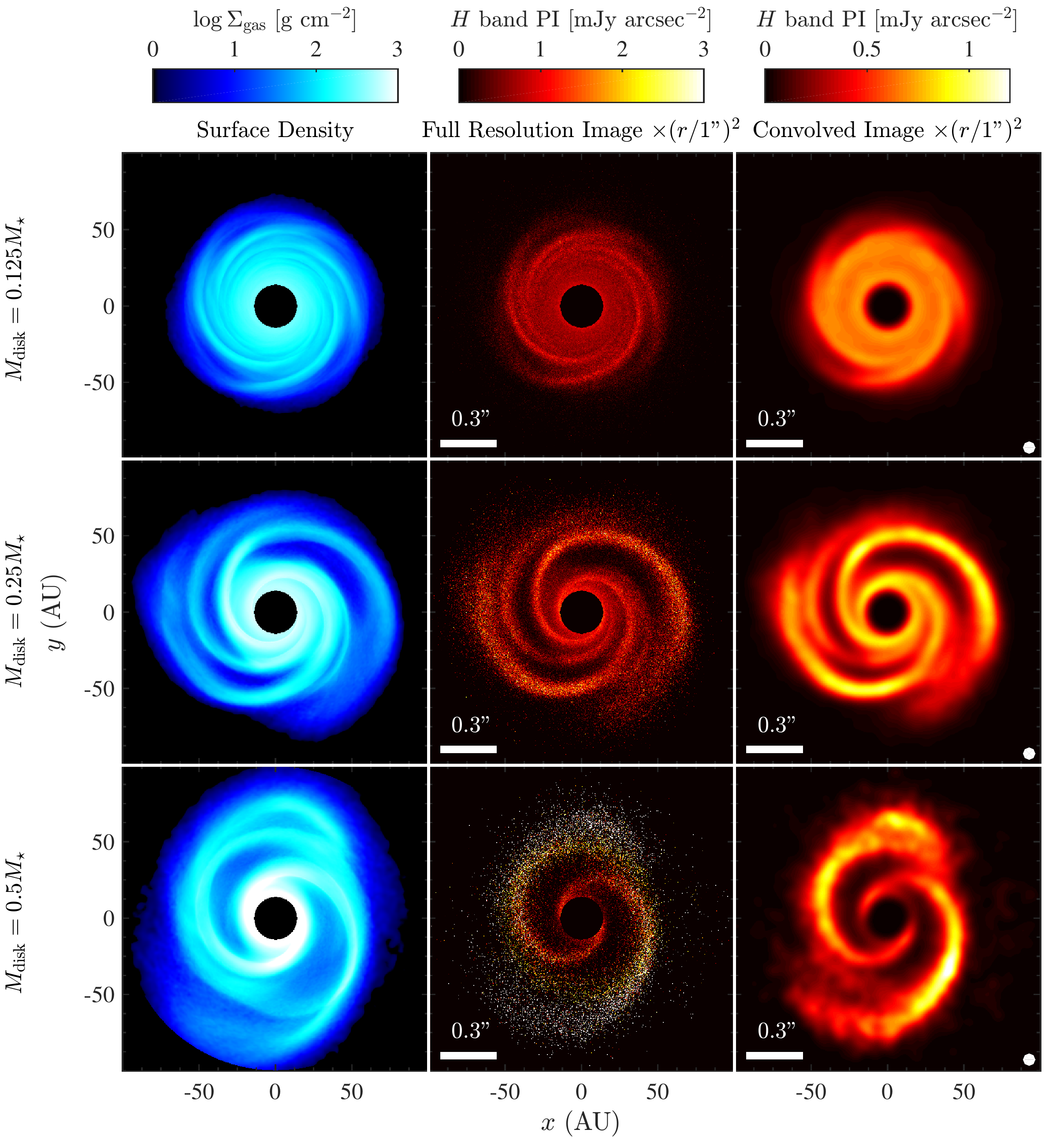}
\end{center}
\figcaption{Surface density and $H$ band PI images at face-on viewing angles (scaled by $r^2$) and at $\sim$1800~yr of the models at 140 pc. Systems are assumed to be at 140 pc. Convolved images (right column) are full resolution MCRT images (middle column) convolved by a Gaussian PSF with a FWHM of $0.05\arcsec$ (marked at the lower right corner). The inner 14 AU ($0.1\arcsec$) in all panels is masked out.
\label{fig:image}}
\end{figure}

\begin{figure}
\begin{center}
\epsscale{1} \plotone{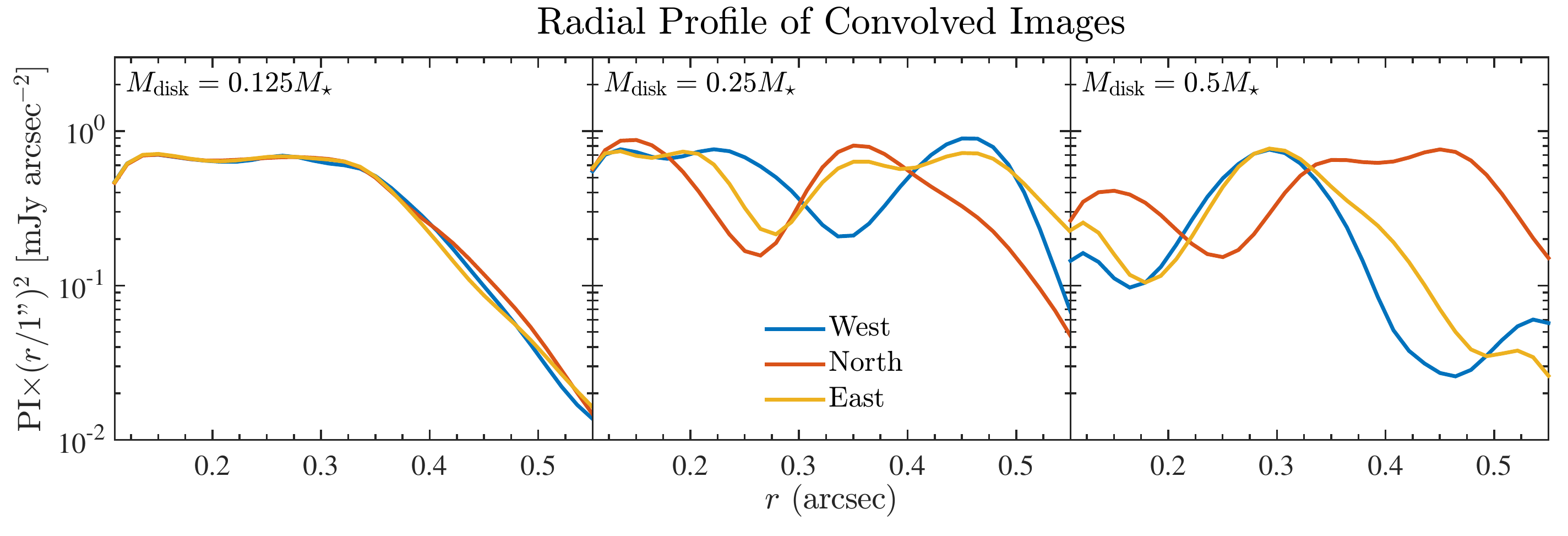}
\epsscale{1} \plotone{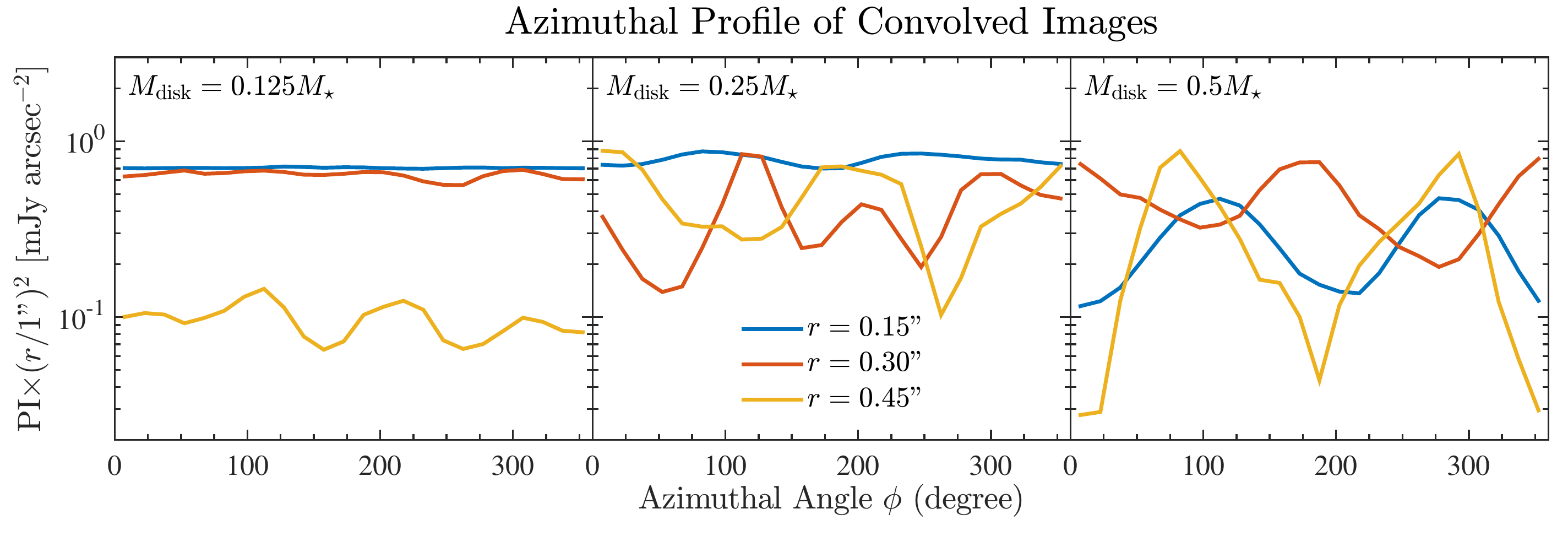}
\end{center}
\figcaption{Radial profiles in the east, north, and west directions (top) and azimuthal profiles at $r=0.15\arcsec$, 0.3$\arcsec$, and 0.45$\arcsec$ (bottom) of the $r^2$-scaled convolved images in Figure~\ref{fig:image}. The peaks on all curves indicate interceptions with the spiral arms. In the azimuthal profiles, the double-peak pattern (with the 2 peaks separated by $\sim$180$^\circ$) is prominent in MD05 at all radii, indicating the dominance of the $m=2$ mode; the quadrupole-peak pattern (with the peaks separated by $\sim$90$^\circ$) is present but less prominent in MD025; the MD0125 model only shows a quadrupole-peak pattern at large radius and low amplitude arms at smaller radius.
\label{fig:profile}}
\end{figure}

The surface density and synthetic $H$ band images are shown in Figure~\ref{fig:image}, and the radial as well as azimuthal profiles of the convolved images are shown in Figure~\ref{fig:profile}. In $\sigmagas$, spiral density waves are clearly present in all cases, while their amplitudes increase from a factor of $\sim$1.5 in MD0125 to a factor of $\sim$3 in MD05 ($\Sigma_{\rm peak}/\Sigma_{\rm b}$ in Table~\ref{tab:models}). In the convolved images, only MD025 and MD05 show prominent spiral patterns, while the arms in MD0125 are weak and hard to distinguish from the background. In particular, the MD05 model shows a pair of grand-design $m=2$ arms, with a morphology very similar to the pair in the MWC 758 system. The contrast of the arms to the background ($I_{\rm peak}/I_{\rm b}$ in Table~\ref{tab:models}) ranges from a factor of $\sim$2 in MD025 to a factor of $\sim$3 in MD05. Note that our arm contrast in convolved images should be taken as upper limits, as realistic instrumental effects such as flux loss and observational noise, which may affect both the absolute and relative fluxes, are not included in the production of our model images (beyond the intrinsic Monte Carlo noise in the radiative transfer images). The pitch angle of the arms (Table~\ref{tab:models}) increases from $10^\circ$ in MD0125 to $15^\circ$ in MD05, values that are consistent with those characterizing the SAO 206462 and MWC 758 systems \citep{dong15-spiralarms}.


\section{Summary and Connection with Observations}\label{sec:summary}

\begin{figure}
\begin{center}
\epsscale{0.7} \plotone{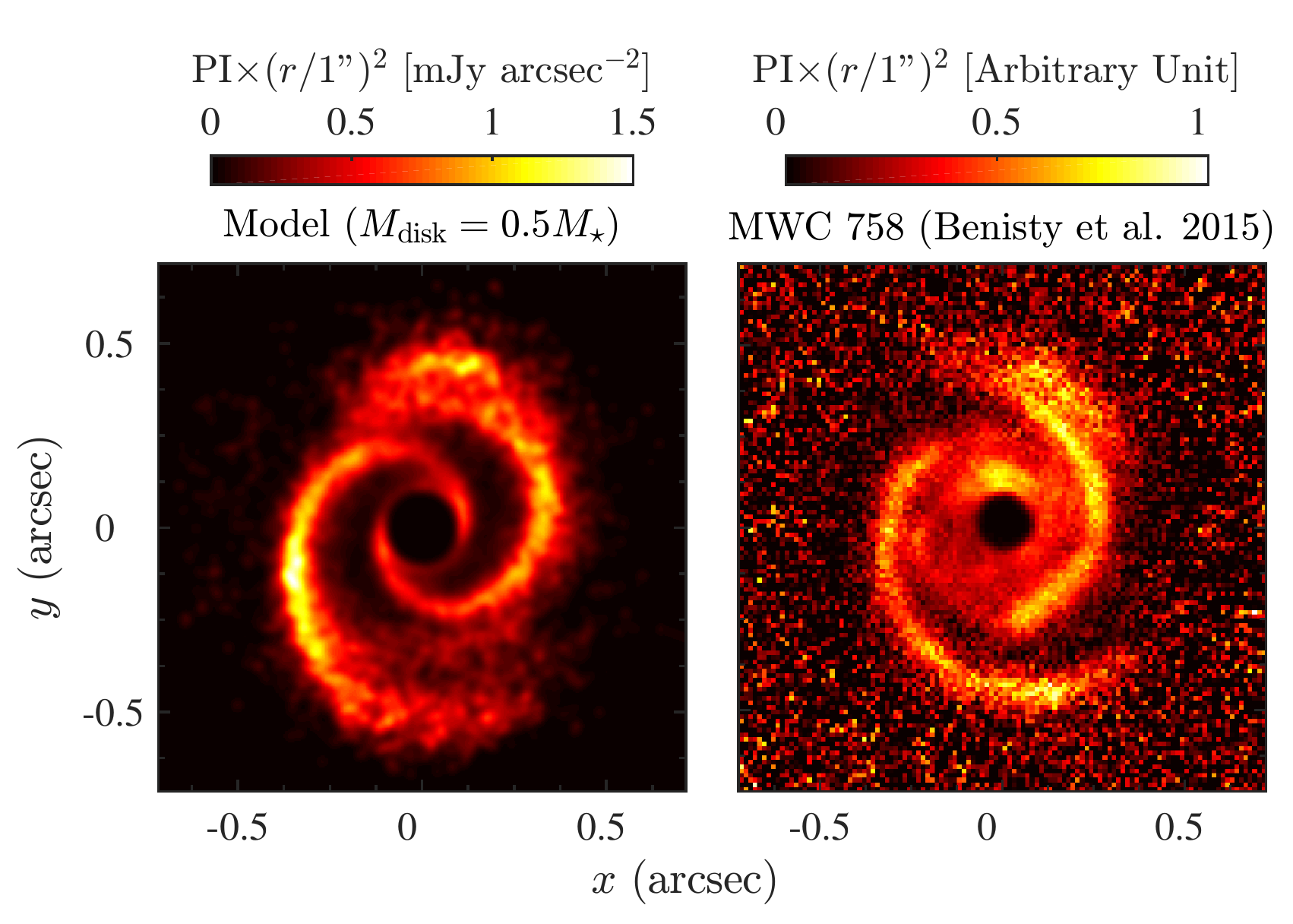}
\epsscale{0.7} \plotone{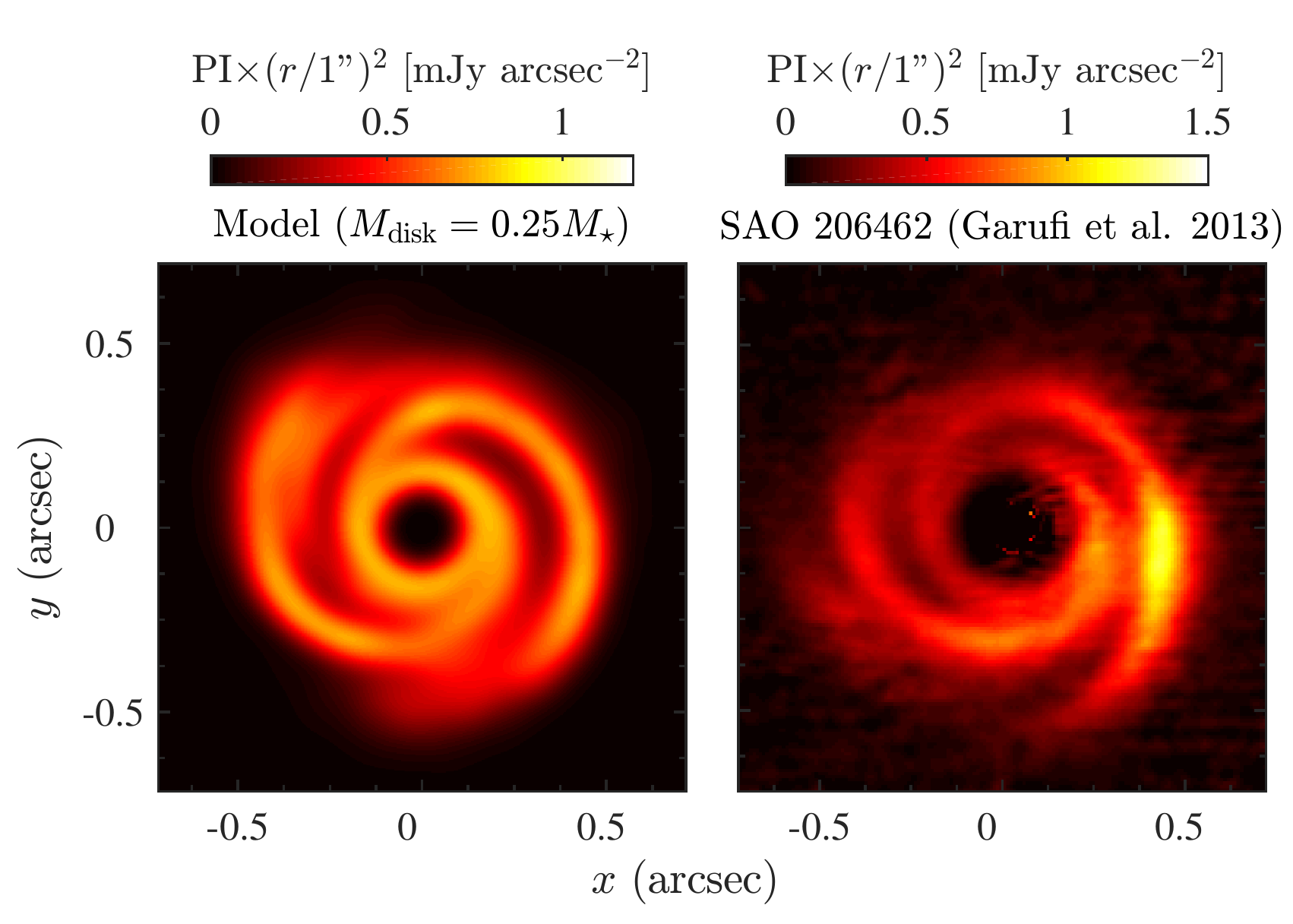}
\end{center}
\figcaption{Comparisons of the $q=0.5$ model with MWC 758 \citep[][top]{benisty15}, and the $q=0.25$ model with SAO 206462 \citep[][bottom]{garufi13}. The models are assumed to be at 140 pc, and convolved by a Gaussian PSF with FWHM=$0.03\arcsec$ (MD05) and FWHM=$0.09\arcsec$ (MD025) to match the corresponding observations. The units in all images are mJy arcsec$^{-2}$, except for MWC 758 for which the units are arbitrary.
\label{fig:comparison}}
\end{figure}

We studied the observational signatures of GI-induced spiral arms in direct imaging observations at NIR wavelengths with 3D SPH calculations and 3D MCRT simulations. Our main conclusions are:
\begin{enumerate}
\item Gravitational instability can produce prominent spiral arms in NIR scattered light images when $q=\mdisk/\mstar\gtrsim0.25$. Their morphologies (see Figure \ref{fig:comparison}) resemble strongly the ones discovered in recent observations (e.g., SAO 206462 as imaged by \citealt{garufi13},
and most strikingly, the pair of $m=2$ arms 
in MWC 758 found by \citealt{benisty15}).
\item When $q\lesssim0.1$, GI-induced spiral arms are
small in size and weak in scattered light images, with contrasts $\lesssim1.5$. They may be difficult to detect with the current angular resolution of NIR observations. A similar conclusion was reached by \citet{dipierro15-gidisks}.
\item The dominant azimuthal wavenumber of GI-induced spiral arms is $m\sim1/q$ in the range of $2\lesssim1/q\lesssim8$.
\item The spiral arms have a pattern speed that roughly agrees with the Keplerian speed at where the arms are. Dust trapping by arms at local pressure maxima may therefore be possible.
\end{enumerate}

Although qualitatively our $q=0.5$ and $q=0.25$ model images show good agreement with the MWC 758 and SAO 206462 observed images (Figure~\ref{fig:comparison}), whether GI is actually at work in these two systems is unclear. There are at least 3 potential issues that need to be addressed to validate GI as a possible explanation for these systems:
\begin{enumerate}
\item Our set of 3 models shows that $q$ has to be $\gtrsim 0.25$ for the spiral density waves to be prominent in scattered light images as seen in observations (in particular, grand-design $m=2$ arms require $q\approx0.5$). However, under conventional assumptions for dust opacity and dust-to-gas-mass ratio, \citet{andrews11} estimated the disk mass $\mdisk$ of these 2 systems using mm continuum observations, and concluded that $q$ is on the order of 1\% in both cases. The caveat, of course, is that the conventional assumptions for dust opacity and dust-to-gas-mass ratio may not be correct, and $\mdisk$ may have been underestimated, at least in some cases. See our discussion on the HL Tau system below. On the other hand, disks with $q\gtrsim0.25$ may develop substantial non-Keplerian velocity components, the detections of which will help determine the masses of these systems \citep{rosenfeld14}.

\item The measured stellar
mass accretion rates of the 2 systems are both on the order of $10^{-8}\msun$ yr$^{-1}$ \citep{eisner09,grady09}, about 2 orders of magnitude smaller than the $\mdot$'s from our models MD025 and MD05. The caveat here is that accretion in GI disks can be highly episodic. As shown in \citet{vorobyov15} (their Figure~3), although the time-averaged $\mdot$ can reach $\sim10^{-6}\msun$ yr$^{-1}$, $\mdot$ fluctuates with large amplitude and can occasionally dip under $10^{-8}\msun$ yr$^{-1}$.

\item While the spiral arms in SAO 206462 are located at $\sim$30--80 AU from the star, consistent with our models, the $m=2$ arms in MWC 758 extend to $\sim$135~AU, which is formally outside the region that
we simulate
(MWC 758 is 280 pc away, \citealt{vanleeuwen07}; SAO 206462 is 140 pc away, \citealt{muller11}; our model distances are 140 pc). GI disks tend to fragment outside a critical radius $r_{\rm f}$ if the opacity has the form of $\kappa\propto T^2$, as expected from the ice-grain-dominated opacity at low temperatures \citep{rafikov05, rice09, forgan11, forgan13}. Both theoretical estimates \citep{rafikov09, clarke09, kratter10} and some numerical simulations \citep[e.g.][]{stamatellos07} have suggested $r_{\rm f}$ to be around $\sim$100~AU (but see also Figure 2 in \citet{vorobyov13} and Figure 1 in \citet{vorobyov15}, showing spiral arms but no fragments at several hundred AU). The radial scale of MWC~758's arms appear dangerously close to, if not exceeding, the fragmentation radius.
\end{enumerate}

We conclude from the above three concerns that it is unclear whether the major spiral arm systems discovered so far from direct NIR imaging surveys can be attributed to GI. Planets, as shown by \citet{dong15-spiralarms}, represent a more plausible explanation. Nevertheless, if future NIR imaging observations find prominent spiral 
arms in disks (1) with large masses
($q \gtrsim 0.25$), (2) with high
stellar accretion rates 
($\mdot\gtrsim10^{-6}\msun$ yr$^{-1}$),
and (3) on relatively compact scales ($\lesssim$ 100~AU, safely inside
the gravitational fragmentation radius), then GI would be securely implicated. Once such systems are found, the number of spirals in the disk, i.e., the dominant azimuthal wavenumber, will be a good diagnostic of the disk's total mass.

We close with a few remarks on the HL Tau system. With a high accretion rate ($9\times10^{-6}\msun$ yr$^{-1}$, \citealt{hayashi93}) and total gas disk mass estimated from mm continuum observations ($q\sim0.25$, \citealt{kwon11}), the HL Tau system should be prone to gravitational instability, as suggested by our model MD025, and prominent low $m$-mode spiral arms are expected to form. This picture seems to be incompatible with the ALMA observations \citep{brogan15} that reveal concentric rings, likely created by planets \citep{dong15-gaps, dipierro15-hltau}. The lack of spirals may suggest that previous estimates of HL Tau's disk mass are too high and/or that the stellar mass estimate is too low.


\section*{Acknowledgments}
R.D. thanks Cathie Clarke and Roman Rafikov for educating him on the subject of GI. We thank Myriam Benisty and Antonio Garufi for kindly sharing with us the VLT/SPHERE image of MWC 758, and the VLT/NACO image of SAO 206462, respectively. We are grateful to the anonymous referee for constructive suggestions that improved the quality of the paper. This project is partially supported by NASA through Hubble Fellowship grant HST-HF-51320.01-A (R.D.) awarded by the Space Telescope Science Institute, which is operated by the Association of Universities for Research in Astronomy, Inc., for NASA, under contract NAS 5-26555. E.C. acknowledges support from
NASA and the NSF. Numerical calculations were performed on the SAVIO cluster provided by the Berkeley Research Computing program, supported by the UC Berkeley Vice Chancellor 
for Research and the Berkeley Center for Integrative Planetary Science.


\clearpage

\begin{table}
\tiny
\begin{center}
\caption{Model Properties}
\begin{tabular}{ccccccccc}
\tableline\tableline
Model & $q=\mdisk/\mstar$ & $\mdot$ & Dominant Azimuthal Number & $\omegap$\tablenotemark{a} & $r_{\rm CR}$\tablenotemark{b} & $\Sigma_{\rm peak}/\Sigma_{\rm b}$\tablenotemark{c} & $I_{\rm peak}/I_{\rm b}$\tablenotemark{d}  & Pitch Angle\tablenotemark{e} \\
 &  & ($\msun$ yr$^{-1}$) & & & (AU) & & & (degree) \\
\tableline
MD0125 & 0.125 & $7\times10^{-8}$ & $m\sim4-8$ & $\sim2\pi/400$ yr & 56 & [1.34, 1.60] & [1.06, 1.50] & 10 \\
MD025 & 0.25 & $5\times10^{-7}$ & $m\sim4$ & $\sim2\pi/310$ yr  & 49 & [3.48, 2.28] & [2.10, 1.83] & 11 \\
MD05 & 0.5 & $5\times10^{-6}$ & $m=2$ & $\sim2\pi/280$ yr  & 47 & [2.50, 3.20] & [1.71, 2.87] & 15 \\
\tableline
\end{tabular}
\tablecomments{(a) Pattern speed of the spiral density waves. As the waves are not in exact rigid body rotation, $\omegap$ is evaluated during a segment of the simulation when the  morphologies of the waves did not appear to evolve much by eye. (b) Corotation radius at which $\Omega_{\rm Keplerian}(r_{\rm CR})=\omegap$. (c) Surface density amplitude of the waves evaluated
at $\sim$1800~yr 
and at [$r=0.3\arcsec$, $r=0.45\arcsec$] ([$r$=42~AU, $r$=63~AU] for an assumed distance of 140 pc). $\Sigma_{\rm peak}$ is the peak surface density; $\Sigma_{\rm b}$ is the background surface density taken to be the azimuthal average at the same radius. (d) Same as (c), but for scattered light intensity in convolved images. (e) Averaged pitch angle of the arms in convolved images.}
\label{tab:models}
\end{center}
\end{table}

\end{document}